\documentclass[a4paper,11pt]{article}
\usepackage{pos}

\usepackage{bm}
\let\vec=\bm
\newcommand{\cP}{{\cal P}}
\newcommand{\cK}{{\cal K}}
\renewcommand\Re{\operatorname{Re}}

\title{Full solution of the medium-induced radiation spectrum}

\author*[a,b]{Carlota Andres}
\author[b,c]{Liliana Apolin\'ario}
\author[d]{Fabio Dominguez}

\affiliation[a]{Jefferson Lab,\\
 Newport News, Virginia 23606, US}

\affiliation[b]{LIP, \\
Av. Prof. Gama Pinto, 2, P-1649-003 Lisboa, Portugal}

\affiliation[c]{Instituto Superior T\'{e}cnico (IST), Universidade de Lisboa,\\
 Avenida Rovisco Pais 1, 1049-001 Lisbon, Portugal}
 
\affiliation[d]{Instituto Galego de F\'isica de Altas Enerx\'ias IGFAE, Universidade de Santiago de Compostela,\\
 E-15782 Santiago de Compostela (Galicia-Spain)}

\emailAdd{carlota@jlab.org}
\emailAdd{liliana@lip.pt}
\emailAdd{fabio.dominguez@usc.es}

\abstract{New measurements of jet quenching observables at RHIC and at the LHC, such as jet substructure observables, demand an increased precision in the theory calculations describing medium-induced radiation of gluons. Closed expressions for the gluon spectrum including a full resummation of multiple scatterings have been known for the past 20 years. Still they have only been evaluated in specific limits either taking a few terms in an opacity expansion or by employing a Gaussian approximation for the interaction potential -- which misses essential physical effects. We present here a new flexible method to compute the full spectrum for a realistic interaction potential, thus allowing us for the first time to properly quantify the effect of the all-order resummation of multiple scatterings. This new approach paves the way for precision phenomenological studies including multiple scattering effects such as coherence phenomena.}

\FullConference{%
  HardProbes2020\\
  1-6 June 2020\\
  Austin, Texas}

\begin{document}
\maketitle

\section{Introduction and derivation}

New measurements of hard probes observables in high-energy heavy-ion collisions, together with recent studies suggesting that jet quenching observables may provide insight into the different stages of the evolution of the system created after the collision \cite{Apolinario:2017sob,Andres:2019eus}, make imperative to revisit and relax approximations under which the in-medium radiation spectrum is computed. The medium-induced soft gluon spectrum off a high-energy parton in the BDMPS-Z  framework reads:
\begin{equation}
\omega\frac{dI}{d\omega d^2\vec{k}} = \frac{2\alpha_s C_R}{(2\pi)^2\omega^2} \Re \int_0^\infty dt' \, \int_0^{t'} dt \,\int_{\vec{p}\vec{q}} 
 \vec{p} \cdot \vec{q} \,\,\widetilde{\cK}(t',\vec{q};t,\vec{p})  \cP(\infty,\vec{k};t',\vec{q})\:,
 \label{eq:bdmps}
\end{equation}
where $\vec{k}$ and $\omega$  are, respectively, the two-dimensional transverse momentum and energy of the emitted gluon\footnote {For further details about Eq.~(\ref{eq:bdmps}) and the explicit expressions of the kernel  $\widetilde\cK (t',\vec{q};t,\vec{p})$ and the momentum broadening factor $\cP(\infty,\vec{k};t',\vec{q})$ we refer the reader to \cite{Andres:2020vxs}.}. Since the numerical evaluation of the kernel $\widetilde\cK (t',\vec{q};t,\vec{p})$, including all the multiple scatterings for a realistic collision rate $V(\vec{q})$ -- such as a Yukawa interaction -- is challenging, the spectrum in Eq.~(\ref{eq:bdmps}) has been historically computed within two approximations in which an analytical expression for the kernel is possible: multiple soft and single hard momentum transfer. The former, also known as the \emph{harmonic oscillator} (HO) approximation, assumes a Gaussian profile for the transverse momentum transfer. The latter is obtained by expanding the integrand of the kernel in powers of the density of scattering centers (\emph{opacity expansion}). The first order ($N=1$) in this procedure is usually referred to as the Gyulassy-Levai-Vitev (GLV) approximation.

 Following the direction of~\cite{CaronHuot:2010bp}, we present here a method  that allows us to compute the full spectrum, including the resummation of all multiple scatterings, for realistic interactions without the need of using the above mentioned approximations. We start by writing the spectrum as
\begin{equation}
\omega\frac{dI}{d\omega d^2\vec{k}} = \frac{2\alpha_s C_R}{(2\pi)^2\omega} \Re\int_0^L ds\; n(s) \int_0^s dt\int_{\vec{p}\vec{q}\vec{l}}i\vec{p}\cdot\left(\frac{\vec{l}}{\vec{l}^2}-\frac{\vec{q}}{\vec{q}^2}\right)\sigma(\vec{l}-\vec{q})\widetilde\cK(s,\vec{q};t,\vec{p})\cP(L,\vec{k};s,\vec{l})\:,
\label{eq:fullspect}
\end{equation}
which is equivalent to  Eq.~(\ref{eq:bdmps}) once the vacuum contribution has been subtracted (this is shown in Appendix~A of ref.\cite{Andres:2020vxs}). The broadening satisfies
\begin{equation}
\partial_\tau \cP(\tau,\vec{k};s,\vec{l}) = -\frac{1}{2}n(\tau)\int_{\vec{k}'}\, \sigma(\vec{k}-\vec{k}')\cP(\tau,\vec{k}';s,\vec{l})\:,
\end{equation}
with initial condition $\cP(s,\vec{k};s,\vec{l})=(2\pi)^2\delta^{(2)}(\vec{k}-\vec{l})$ and the kernel satisfies
\begin{equation}
\partial_{t}\widetilde{\cK}(s,\vec{q};t,\vec{p}) = \frac{i\vec{p}^2}{2\omega}\widetilde{\cK}(s,\vec{q};t,\vec{p})+\frac{1}{2}n(t)\int_{\vec{k}'}\sigma(\vec{k}'-\vec{p})\widetilde{\cK}(s,\vec{q};t,\vec{k}')\:,
\label{eq:diffKt}
\end{equation}
with initial condition $\widetilde\cK(s,\vec{q};s,\vec{p})=(2\pi)^2\delta^{(2)}(\vec{q}-\vec{p})$; where $\sigma$ is the dipole cross section which can be written, in momentum space,  in terms of the collision rate $V$ as
\begin{equation}
\sigma(\vec{q}) = -V(\vec{q}) + (2\pi)^2 \delta^2(\vec{q}) \int_{\vec{l}} V(\vec{l})\,.
\label{eq:dipole_xsection}
\end{equation}

Instead of trying to solve the previous differential equations, we define 
\begin{equation}
\vec{\phi}(\tau,\vec{k};s,\vec{q}) = n(s)\int_{\vec{l}}\left(\frac{\vec{l}}{\vec{l}^2}-\frac{\vec{q}}{\vec{q}^2}\right)\sigma(\vec{l}-\vec{q})\cP(\tau,\vec{k};s,\vec{l})\:,
\end{equation}
which satisifies
\begin{equation}
\partial_\tau \vec{\phi}(\tau,\vec{k};s,\vec{q}) = -\frac{1}{2}n(\tau)\int_{\vec{k}'}\, \sigma(\vec{k}-\vec{k}')\vec{\phi}(\tau,\vec{k}';s,\vec{q})\:,
\label{eq:diffphi}
\end{equation}
with initial condition
\begin{equation}
\vec{\phi}(s,\vec{k};s,\vec{q}) = n(s)\left(\frac{\vec{k}}{\vec{k}^2}-\frac{\vec{q}}{\vec{q}^2}\right)\sigma(\vec{k}-\vec{q})\:;
\label{eq:initphi}
\end{equation}
and
\begin{equation}
\vec{\psi}_I(s,\vec{k};t,\vec{p}) = e^{\frac{i\vec{p}^2}{2\omega}(s-t)}\,
 \int_{\vec{q}}\vec{\phi}(L,\vec{k};s,\vec{q}) \,\widetilde\cK(s,\vec{q};t,\vec{p}) \:,
\end{equation}
satisfying
\begin{equation}
\partial_{t}\vec{\psi}_I(s,\vec{k};t,\vec{p}) = \frac{1}{2}n(t)\int_{\vec{k}'}e^{\frac{i\vec{p}^2}{2\omega}(s-t)}\sigma(\vec{k}'-\vec{p})e^{-\frac{i\vec{k}^{\prime2}}{2\omega}(s-t)}\vec{\psi}_I(s,\vec{k};t,\vec{k}')\:,
\label{eq:diffpsiI}
\end{equation}
with initial condition $\vec{\psi}_I(s,\vec{k};s,\vec{p}) = \vec{\phi}(L,\vec{k};s,\vec{p})\:.$

Then, the full $\vec{k}$-dependent spectrum can then be written as
\begin{equation}
\omega\frac{dI}{d\omega d^2\vec{k}} = \frac{2\alpha_s C_R}{(2\pi)^2\omega} \Re\int_0^L ds \int_0^s dt\int_{\vec{p}}ie^{-\frac{i\vec{p}^2}{2\omega}(s-t)}\vec{p}\cdot\vec{\psi}_I(s,\vec{k};t,\vec{p})\:.
\label{eg:specpsiI}
\end{equation}
One can integrate over transverse momentum the previous expression to obtain the gluon energy distribution, but always keeping in mind that the integration must respect the kinematical constraint $k < \omega$ since the derivation of the medium-induced emission spectrum assumes that the transverse momentum of the radiated gluon is small ($k \ll \omega$).

In order to evaluate this spectrum, first we solve numerically the differential equation satisfied by $\phi$, Eqs.~({\ref{eq:diffphi}) and~(\ref{eq:initphi}), which is used to compute the initial condition for that of $\vec{\psi}_I$. Now we can obtain the solution for $\vec{\psi}_I$ by solving its differential equation. Once we have a solution for $\vec{\psi}_I$, we plug it into Eq.~(\ref{eg:specpsiI}) and integrate numerically to obtain the $\vec{k}$-dependent
spectrum. This can be done for different realistic interaction models, such as the Yukawa and hard thermal loop parton-medium interactions, which enter in the previous equations through the collision rate $V$ defined in Eq.~(\ref{eq:dipole_xsection}).

 \section{Numerical results}
 
In this section we present the results of our numerical analysis. For simplicity, we assume a constant linear density of scatterings $n(t)=n_0$ and we take $C_R=4/3$ and $\alpha_s=0.3$. The results of our approach are presented here for a Yukawa-type interaction given by $V(\vec{q}) = 8\pi\mu^2/(\vec{q}^2+\mu^2)^2$\footnote{For the results obtained using the hard thermal loop (HTL) formalism \cite{Aurenche:2002pd} see ref.~\cite{Andres:2020vxs}.}.

We present in Fig.~\ref{fig:spec_yukawa_severalL} a comparison between the full resummed medium-induced gluon energy distribution (solid lines) and the GLV $N=1$ approximation (dash-dotted lines) for $n_0 = 1 \,\mathrm{fm^{-1}}$ and $\mu = 0.6$~GeV while varying the medium path length $L$\footnote{The expression of GLV $N=1$ spectrum can be found in Appendix~B of ref.~\cite{Andres:2020vxs}.}. For a fixed linear density, the larger the path length the larger the discrepancy between the full resummed and GLV $N=1$ results since the opacity expansion is only well justified for small values of $n_0L$.  Moreover, at large gluon energies ($\omega > \bar\omega_c \equiv \mu^2L/2$), the magnitudes of the full resummed and  GLV spectra are, as expected, very similar, since in this kinematical region the interaction is believed to be dominated by a single hard scattering. To illustrate the asymptotic behavior of the two approaches, we use in the right panel of this figure a  logarithmic scale for both axes, which makes clear that both spectra are suppressed as $\bar \omega_c /\omega$ for $\omega > \bar \omega_c$. At smaller energies  ($\omega < \bar\omega_c$) the GLV $N=1$ evaluation is much larger than the full one, given that the former does not account for the coherent effect of multiple scatterings. 

\begin{figure}
\vspace*{-5mm}
\centering
\includegraphics[scale=0.42]{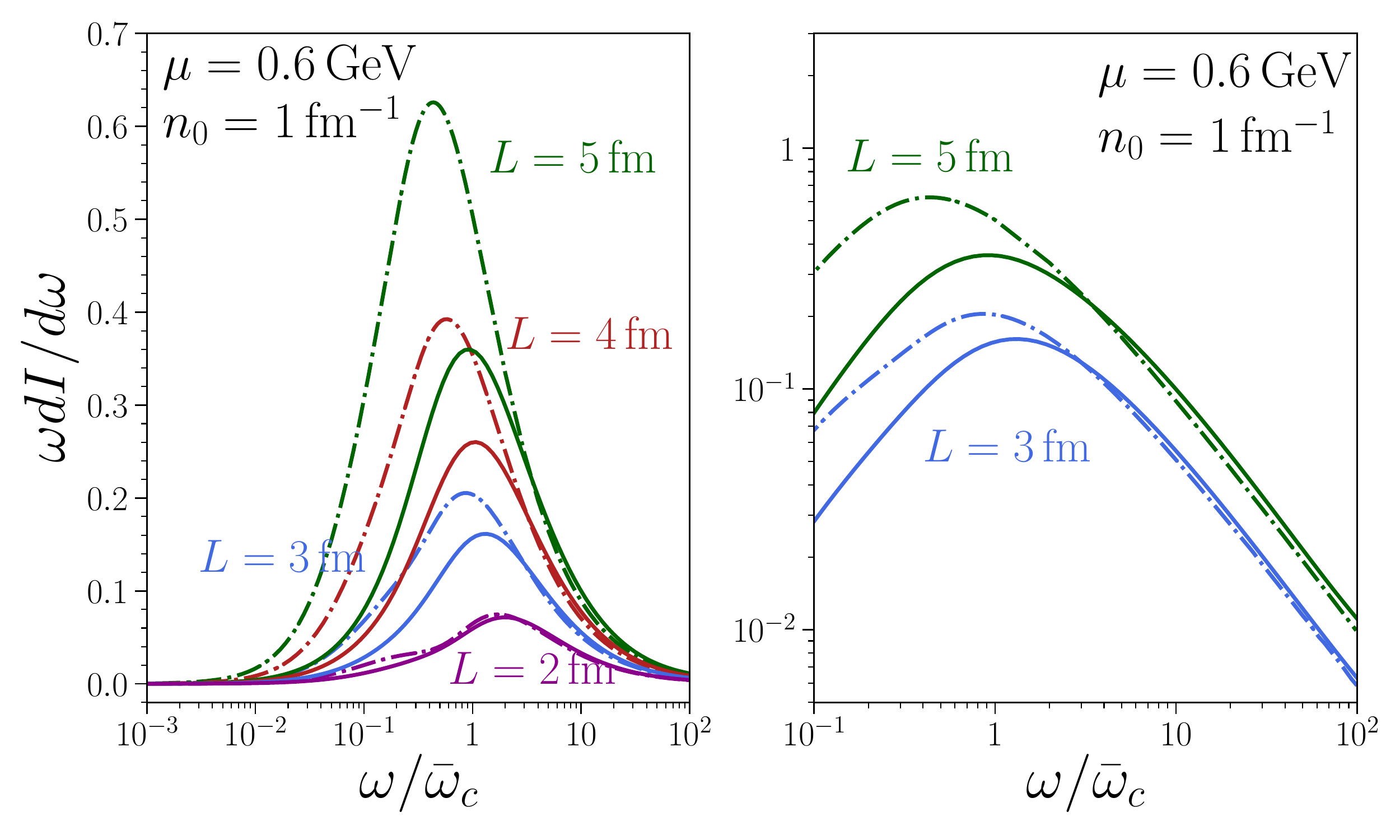}
\vspace*{-2mm}
\caption{Left: full medium-induced gluon energy distribution for the Yukawa collision rate (solid lines) compared to the GLV $N=1$ result (dash-dotted lines) with $\mu=0.6$ GeV and linear density $n_0 = 1 \,\mathrm{fm^{-1}}$ for different values of $L$ as a function of the rescaled gluon energy $\omega/\bar\omega_c \equiv 2\omega/ \mu^2L$. Right: same as left panel for only two values of $L$ in log-log scale.} 
\label{fig:spec_yukawa_severalL}
\end{figure}

A comparison between the full medium-induced gluon energy distribution and the HO result is presented in Fig.~\ref{fig:spec_yukawa_harmonic}. It is worth noticing that this comparison is not straightforward and must be taken with care as these approaches do not involve the same parameters. We will fix here $\hat{q}L = 1.3 \left(n_0 L \right) \mu^2$ in order to make qualitative comparisons between these evaluations. The solid lines  in Fig.~\ref{fig:spec_yukawa_harmonic} represent the full resummed result for $\mu =1.6$ GeV (or $\bar\omega_c=39$ GeV), in blue, and $\mu = 0.9$ GeV (or $\bar\omega_c =12.3$ GeV), in red, and the dotted ones the HO evaluation for the corresponding values of $\hat{q}$ i.e., $\hat{q} = 2.8 \,\mathrm{GeV^2/fm}$ (blue) and $\hat{q} = 0.9\, \mathrm{GeV^2/fm}$ (red). For large gluon energies ($\omega > \bar \omega_c$), contrarily to the GLV evaluation, the HO  calculation disagrees substantially with the full resummed one, since in this kinematic region, dominated by a single hard scattering, the Gaussian approximation is not well justified. In fact, the HO approximation goes much faster to zero in this region -- proportionally to $1/\omega^2$ --  than the full result. For lower energies ($\omega < \bar \omega_c$) the two evaluations still differ. However, in this region the HO calculation is a much better approximation to the full one than the GLV $N=1$ (see Fig.~\ref{fig:spec_yukawa_severalL}).
\begin{figure}[th]
\centering
\vspace*{-5mm}
\includegraphics[scale=0.37]{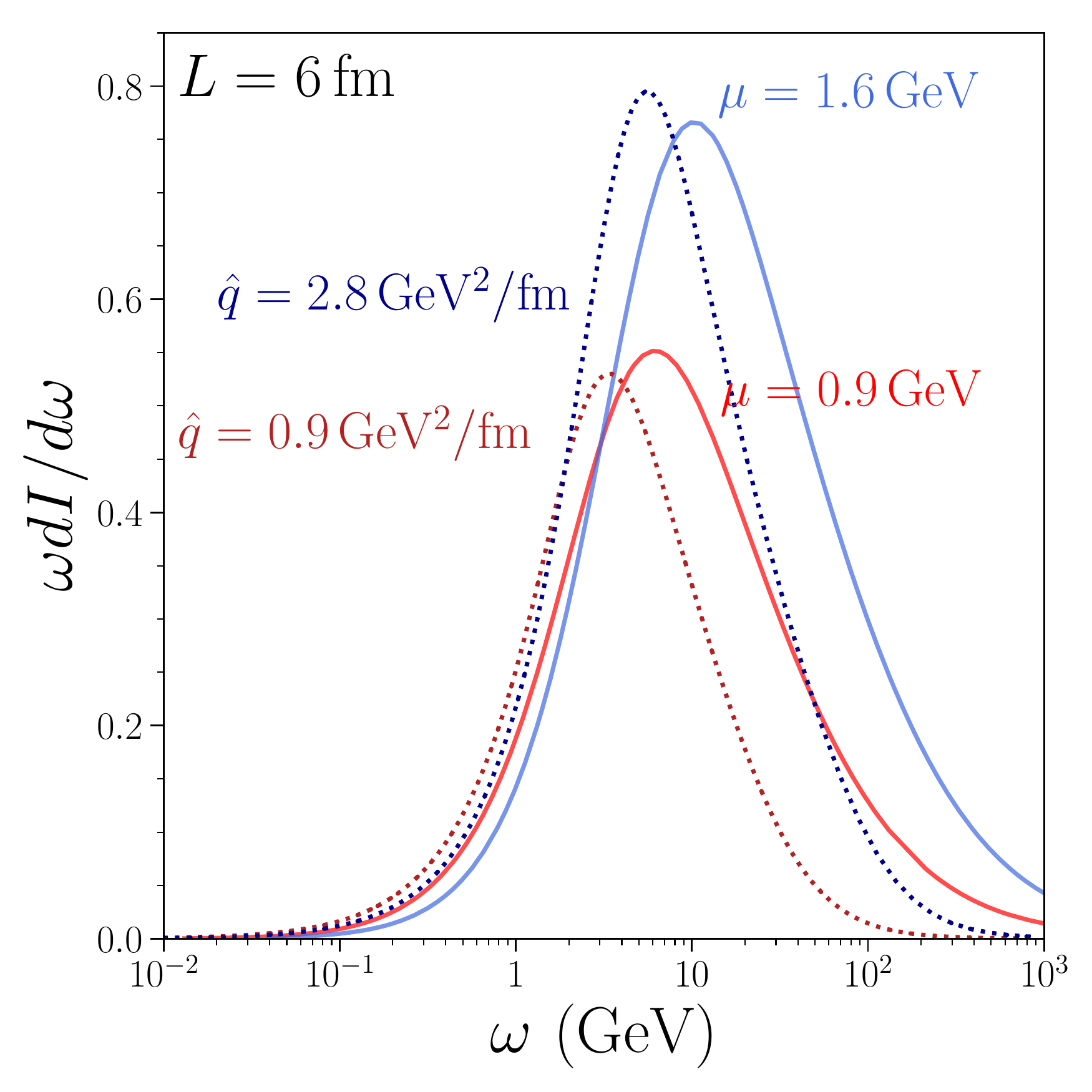}
\vspace*{-2mm}
\caption{Solid lines: full medium-induced gluon energy distribution for the Yukawa collision rate with $\mu=1.6$ GeV (blue) and $\mu=0.9$ GeV (red) for a medium with $n_0L=5$ and $L=6$ fm vs. the gluon energy $\omega$. Dotted lines: medium-induced gluon energy distribution within the HO approximation for $\hat{q} = 2.8 \,\mathrm{GeV^2/fm}$ (blue) and $\hat{q} = 0.9\, \mathrm{GeV^2/fm}$ (red) and $L=6$ fm vs. $\omega$.} 
\label{fig:spec_yukawa_harmonic}
\end{figure}

 \section{Conclusions}
 
We present here a procedure to evaluate the medium-induced gluon radiation spectrum including the full resummation of multiple scatterings which can be used for any realistic interaction potentials without any further assumptions. This method consists of a set of differential equations that can be easily solved and which allows us to obtain robust results beyond the usually employed HO and GLV $N=1$ approximations.

We compute the full resummed gluon energy distribution for a Yukawa-like interaction and compare it to the results obtained within the HO and GLV $N=1$ approximations, finding the discrepancies among them significant. Particularly, the full resummed calculation is smaller than the GLV $N=1$ approximation for low gluon energies since the latter does not account for coherent effects among multiple scatterings. Nonetheless,  at higher energies ($\omega > \bar \omega_c$), the GLV approximation agrees with the full resummed result, since the interaction is dominated by a single hard scattering in this kinematic region. On the other hand, the HO spectrum decreases much faster than the full one in this region, thus resulting in a softer energy spectrum.  However, at lower gluon energies, the HO calculation is a better approximation to the full evaluation than the GLV $N=1$ result, emphasizing the importance of including the effects of multiple scatterings.

\acknowledgments
CA was supported by the US Department of Energy contract DE-AC05-06OR23177, under which Jefferson Science Associates, LLC operates Jefferson Lab. LA was supported by Funda\c{c}\~{a}o para a Ci\^{e}ncia e Tecnologia (FCT - Portugal) under project DL57/2016/CP1345/CT0004 and CERN/FIS-PAR/0022/2017.

\end{document}